\documentclass[aps,prd,showpacs,twocolumn,amsmath]{revtex4-1}
\usepackage{subfigure}
\usepackage{graphicx}
\usepackage{dcolumn}
\usepackage{bm}
\usepackage{float}
\usepackage{xcolor}
\usepackage{epstopdf}
\usepackage{amsmath}
\usepackage{amsmath}
\usepackage{amssymb}
\definecolor{aa}{RGB}{0,0,139}
\usepackage[bookmarksnumbered=true,colorlinks,urlcolor=aa,linkcolor=blue,anchorcolor=aa,citecolor=aa]{hyperref}
\usepackage[mathlines]{lineno}
\usepackage{multirow}
\usepackage{makecell}
\usepackage{enumerate}
\newcommand{\EE}{e^+e^-}

\newcommand{\jpsi}{J/\psi}

\newcommand*{\ee}{e^+e^-}
\newcommand*{\Sc}{\ensuremath{\Sigma^{+}}}
\newcommand*{\ASc}{\ensuremath{\bar\Sigma^{-}}}
\newcommand*{\eescsc}{\ensuremath{e^{+}e^{-}\to\Sigma^{+}\bar{\Sigma}^{-}}}
\newcommand*{\eegscsc}{\ensuremath{e^{+}e^{-}\to\gamma^{\rm ISR}\Sigma^{+}\bar{\Sigma}^{-}}}
\newcommand*{\eepiscsc}{\ensuremath{e^{+}e^{-}\to\pi^{0}\Sigma^{+}\bar{\Sigma}^{-}}}

\begin{document}

\title{Measurements of $\Sigma$ electromagnetic form factors in the time-like region using the untagged initial-state radiation technique}
\author{M.~Ablikim$^{1}$, M.~N.~Achasov$^{4,b}$, P.~Adlarson$^{75}$, O.~Afedulidis$^{3}$, X.~C.~Ai$^{80}$, R.~Aliberti$^{35}$, A.~Amoroso$^{74A,74C}$, Q.~An$^{71,58}$, Y.~Bai$^{57}$, O.~Bakina$^{36}$
% ...
\\
\vspace{0.2cm}
(BESIII Collaboration)\\
\vspace{0.2cm} {\it
$^{1}$ Institute of High Energy Physics, Beijing 100049, People's Republic of China\\
% ...
$^{3}$ Bochum  Ruhr-University, D-44780 Bochum, Germany\\
$^{4}$ Budker Institute of Nuclear Physics SB RAS (BINP), Novosibirsk 630090, Russia\\
% ...
$^{35}$ Johannes Gutenberg University of Mainz, Johann-Joachim-Becher-Weg 45, D-55099 Mainz, Germany\\
$^{36}$ Joint Institute for Nuclear Research, 141980 Dubna, Moscow region, Russia\\
% ...
$^{57}$ Southeast University, Nanjing 211100, People's Republic of China\\
$^{58}$ State Key Laboratory of Particle Detection and Electronics, Beijing 100049, Hefei 230026, People's Republic of China\\
% ...
$^{71}$ University of Science and Technology of China, Hefei 230026, People's Republic of China\\
% ...
$^{74}$ University of Turin and INFN, (A)University of Turin, I-10125, Turin, Italy; (B)University of Eastern Piedmont, I-15121, Alessandria, Italy; (C)INFN, I-10125, Turin, Italy\\
$^{75}$ Uppsala University, Box 516, SE-75120 Uppsala, Sweden\\
% ...
$^{80}$ Zhengzhou University, Zhengzhou 450001, People's Republic of China\\
\vspace{0.2cm}
% ...
$^{b}$ Also at the Novosibirsk State University, Novosibirsk, 630090, Russia\\
% ...
}}

\date{\today}

\begin{abstract}
The process $\eescsc$~is studied from threshold up to 3.04~GeV/$c^2$ via the initial-state radiation technique using data with an integrated luminosity of 12.0~fb$^{-1}$, collected at center-of-mass energies between 3.773 and 4.258~GeV with the BESIII detector at the BEPCII collider.
The pair production cross sections and the effective form factors of $\Sigma$ are measured in eleven \Sc\ASc~invariant mass intervals from threshold to 3.04~GeV/$c^2$.
The results are consistent with the previous results from Belle and BESIII. Furthermore, the branching fractions of the decays $J/\psi\to\Sc\ASc$ and $\psi(3686)\to\Sc\ASc$ are determined and the obtained results are consistent with the previous results of BESIII.
\end{abstract}

\maketitle
\section{Introduction}
The inner structure of baryons can be parameterized using electromagnetic form factors (EMFFs). For baryons with spin 1/2, assuming a vector-like current, there are two EMFFs, the magnetic $|G_{\rm M}|$ and the electric $|G_{\rm E}|$ form factors. Experimentally, these can be accessed in the space-like region by electron-baryon elastic scattering and in the time-like region by baryon pair production in electron-positron annihilation~\cite{Geng:2008mf,Brodsky:1974vy,Green:2014xba}.
Despite the fact that much work has been done on the EM structures of protons in both the space-like and time-like regions~\cite{Qattan:2004ht,Puckett:2011xg,CLEO:2005tiu,BaBar:2013ves,CMD-3:2015fvi,BESIII:2019hdp}, experimental information regarding the EMFFs of hyperons remains limited.

The cross section for the process $\ee\to Y\bar{Y}$~via one-photon exchange, where $Y$ denotes a hyperon with spin 1/2, can be expressed in terms of $|G_{\rm E}|$ and $|G_{\rm M}|$~\cite{Cabibbo:1961sz}:
\begin{equation} \label{eq:a00}
\small{\sigma_{Y\bar{Y}}(s)=\frac{4\pi\alpha^{2}C\beta}{3s}\left[|G_{\rm M}(s)|^{2}+\frac{1}{2\tau}|G_{\rm E}(s)|^{2}\right],}
\end{equation}
where $s$ is the square of the center-of-mass (c.m.) energy, $\alpha = 1/137.036$ is the fine-structure constant, $\beta=\sqrt{1-4M_{Y}^{2}/s}$ is the velocity of the final hyperon, $\tau=s/4M_{Y}^2$, and $M_{Y}$ is the mass of the hyperon. The Coulomb correction factor $C$~\cite{Arbuzov:2011ff00,Arbuzov:2011ff}, accounting for the electromagnetic interaction of charged pointlike fermion pairs in the final state, is 1.0 for pairs of neutral hyperons and $y/(1-e^{-y})$ with $y=\pi\alpha(1+\beta^{2})/\beta$ for pairs of charged hyperons. The effective form factor (FF)~\cite{BESIII:2015axk} defined by
\begin{equation} \label{eq:a11}
\small{|G_{\textrm{eff}}(s)|=\sqrt{\frac{2\tau|G_{\rm M}(s)|^{2}+|G_{\rm E}(s)|^{2}}{2\tau+1}}}
\end{equation}
is proportional to the square root of the hyperon pair production cross section.

Experiments have reported cross section measurements for all members of the spin-parity $J^P =(1/2)^{+}$ baryon octet as well as the ground state charmed hyperon $\Lambda_{c}^{+}$ and the $\Omega$ baryon of the 3/2 decuplet~\cite{Huang:2021xte}. Especially close to threshold, intriguing differences are observed.
The cross sections of $\ee\to p\bar{p}$~\cite{BaBar:2013ves,CMD-3:2015fvi,BESIII:2019hdp}, $\ee\to n\bar{n}$~\cite{BESIII:2021np}, $\ee\to\Lambda\bar{\Lambda}$~\cite{BaBar:2007fsu, BESIII:2013vfs,BESIII:2023ioy}, and $\ee\to\Lambda_{c}^{+}\bar{\Lambda}_{c}^{-}$~\cite{BESIII:2017kqg,BESIII:2023kqg} are found to have an abnormal, non-vanishing cross section near threshold. However, a comparably significant effect is not observed for the reactions $\ee\to \Sigma\bar{\Sigma}$~\cite{BESIII:2020uqk, Belle:2022dvb, BaBar:2007fsu, BESIII:2022up}, $\ee\to\Xi\bar{\Xi}$~\cite{BESIII:2021yup, BESIII:2021xup}, and $\ee\to\Omega^{-}\bar{\Omega}^{+}$~\cite{BESIII:2022kzc}.
The unexpected threshold behavior is discussed as final-state interactions~\cite{Dai:2017fwx}, bound states or near threshold meson resonances~\cite{El-Bennich:2008ytt}, or an attractive Coulomb interaction~\cite{Baldini:2007qg,BaldiniFerroli:2010ruh}.

The cross section of the process \eescsc~near its production threshold has been measured by the BESIII experiment~\cite{BESIII:2020uqk} using the scan method. A non-zero threshold cross section was observed with a hint of an enhancement at $\sqrt{s} =$ 2.5 GeV. However, due to limited statistics, an unambiguous conclusion cannot be drawn.

In this paper, a new measurement of the pair production cross sections and the effective form factors of $\Sigma$ from the production threshold up to the invariant mass of $\Sc\ASc$ at 3.04~GeV$/c^2$ with the BESIII detector located at the BEPCII collider is presented. The measurement uses the initial-state radiation (ISR) process $\eegscsc$, where $\gamma^{\rm ISR}$ is a hard photon emitted from the initial $e^+e^-$ pair and thus changes the effective c.m. energy of the collision. The differential cross section for the ISR process is largest when $\gamma^{\rm ISR}$ is emitted almost parallel to the beam axis, where it cannot be detected by BESIII. To benefit from the increased cross section, an untagged ISR measurement is performed.
The differential cross section for the $\eegscsc$ process, integrated over the $\Sc(\ASc)$~momentum and the ISR photon polar angle, reads~\cite{Kuraev:1985hb}:
\begin{equation} \label{ISRLamcs}
\small{\frac{d\sigma_{\eegscsc}\left(q^2\right)}{dq^{2}}=\frac{1}{s}W(s, x)\sigma_{\Sc\ASc}\left(q^{2}\right),}
\end{equation}
where $\sigma_{\Sc\ASc}(q^2)$ is the cross section for the $\eescsc$ process, $q$ is the momentum transfer of the virtual photon, its square equal to the invariant mass squared of $\Sc\ASc$, $x=\frac{2E_{\gamma}^*}{\sqrt{s}}=1-\frac{q^2}{s}$, and $E_{\gamma}^*$ denotes the energy of the ISR photon in the $\EE$ c.m.~system.
The function~\cite{Druzhinin:2011qd}
\begin{equation} \label{corr_ISRfact}
\small{
\begin{aligned}
W(s,x)=kx^{k-1}[1+\frac{\alpha}{\pi}(\frac{\pi^{2}}{2}-\frac{1}{2})+\frac{3}{4}k+k^{2}(\frac{37}{96}-\frac{\pi^{2}}{12}
\\-\frac{1}{72}{\rm ln}\frac{s}{M_{e}})]-k(1-\frac{1}{2}x)+\frac{1}{8}[4(2-x){\rm ln}\frac{1}{x}~~~
\\-\frac{1+3(1-x)^2}{2}{\rm ln}(1-x)-6+x]~~~~~~~~~~~~~
\end{aligned}
}
\end{equation}
describes the probability for the emission of an ISR photon with energy fraction $x$, where $k=\frac{2\alpha}{\pi}[{\rm ln}\frac{s}{M_{e}^2}-1]$ and $M_e$ is the mass of the electron.

\section{BESIII detector and Monte Carlo simulation}
The BESIII detector~\cite{BESIII:2009fln} records symmetric $\EE$ collisions provided by the BEPCII storage ring~\cite{BEPCII}, which operates in the c.m. energy range from 2.0 to 4.95 GeV, with a peak luminosity of $1.0\times10^{33}\rm{cm^{-2}s^{-1}}$ achieved at $\sqrt{s} =$ 3.773 GeV. BESIII has collected large data samples in this energy region~\cite{BESIII:2020nme}. The cylindrical core of the BESIII detector covers 93\% of the full solid angle and consists of a helium-based multilayer drift chamber (MDC), a plastic scintillator time-of-flight system (TOF), and a CsI(Tl) electromagnetic calorimeter (EMC), which are all enclosed in a superconducting solenoidal magnet providing a 1.0 T magnetic field~\cite{Huang:2022wuo}. The solenoid is supported by an octagonal flux-return yoke with resistive plate counter based muon identification modules interleaved with steel. The charged-particle momentum resolution at 1 GeV/$c$ is 0.5\%, and resolution of the specific ionization energy loss ($dE/dx$) in the MDC is 6\% for electrons from Bhabha scattering. The EMC measures photon energies with a resolution of 2.5\% (5\%) at 1 GeV in the barrel (end cap) region. The time resolution in the TOF barrel region is 68 ps, while that in the end cap region used to be 110 ps. The end cap TOF system was upgraded in 2015 using multi-gap resistive plate chamber technology, providing a time resolution of 60 ps~\cite{BESIII:27a,BESIII:28a,Cao:2020ibk}, which benefits 59.6\% of the data used in this analysis.

The data sets used in this analysis are collected at twelve c.m. energies between 3.773 and 4.258 GeV and correspond to a total integrated luminosity of 12.0~fb$^{-1}$~\cite{BESIII:2023ioy}. The individual data sets and the respective luminosities are listed in Table~\ref{datasamples}.
A \textsc{Geant4}-based~\cite{GEANT4:2002zbu} Monte Carlo (MC) simulation package is used to determine the detection efficiency, optimize event selection criteria, and estimate background contributions. For the data processing and analysis, the BESIII Offline Software System~\cite{Asner:2009zza} framework is used. The MC simulated samples of the signal channel ($\eegscsc$) are generated with the \textsc{ConExc} generator~\cite{Ping:2013jka}.
The \textsc{ConExc} generator considers ISR processes using the radiator function at next-to-leading order accuracy including the vacuum polarization.
The cross section line-shape used for the generation of the signal MC samples is taken from Ref.~\cite{BESIII:2020uqk}. The ISR production of vector charmonium states ($\EE\to\gamma^{\rm ISR}\jpsi$, $\gamma^{\rm ISR}\psi(3686)$) is generated with \textsc{BesEvtGen}~\cite{Ping:2008zz} using the \textsc{VECTORISR} model~\cite{VECISR1, VECISR2}. The angular distributions of the $\Sigma$ in $J/\psi\to\Sc\ASc$ and $\psi(3686)\to\Sc\ASc$ decays are modeled according to experimental data~\cite{BESIII:2020fqg}.
Inclusive MC samples at $\sqrt{s}=3.773$ and $4.178$~GeV are used to investigate possible background contamination. They consist of inclusive hadronic processes ($\EE\to q\bar{q}$, $q=u, d, s$) modeled with the \textsc{LUARLW}~\cite{Lund} at $\sqrt{s}=3.773$~GeV and \textsc{KKMC}~\cite{Jadach:1999vf, KKMC1} at $\sqrt{s}=4.178$~GeV. The dominant background channel, \eepiscsc, is generated exclusively using the phase space \textsc{ConExc} generator.

\begin{table}
	\centering
	\caption{The c.m.~energies $\sqrt{s}$ and the integrated luminosities $\mathcal{L}_{\rm int}$ of each data set~\cite{BESIII:2023ioy}. \label{datasamples}}
	\begin{tabular}{cccc}\hline \hline
		$\sqrt{s}$ $($GeV$)$  &~~~~&&$\mathcal{L}_{\rm int}$ $($pb$^{-1})$\\
		\hline
		3.773             &&&2931.80  \\ 
		4.128             &&&401.50   \\ 
		4.157             &&&408.70   \\ 
		4.178             &&&3189.00  \\ 
		4.189             &&&526.70   \\ 
		4.199             &&&526.00   \\ 
		4.209             &&&517.10   \\ 
		4.219             &&&514.60   \\ 
		4.226             &&&1091.74   \\ 
		4.236             &&&530.30   \\ 
		4.244             &&&538.10   \\ 
		4.258             &&&825.74   \\ 
\hline \hline
	\end{tabular}
\end{table}

\section{Event selection and background analysis}
To select the candidates for $\eegscsc$, the decays of $\Sc(\ASc)\to p\pi^{0}(\bar{p}\pi^{0})$ and $\pi^{0}\to\gamma\gamma$ are reconstructed, while the $\gamma^{\rm ISR}$ is not detected.

The number of charged tracks is required to be two with a net charge of zero. These tracks are reconstructed in the MDC and required to have a polar angle $\theta$ within $\vert\!\cos\theta\vert <$ 0.93, where $\theta$ is defined with respect to the z-axis, which is the symmetry axis of the MDC.

Furthermore, the distance of closest approach of each charged track to the interaction point is required within 2 cm in the plane perpendicular to the beam and within 10 cm in the direction along the z-axis.
The two selected tracks are identified as one proton and one anti-proton by requiring $\mathcal{P}(p)>\mathcal{P}(h)$, where $\mathcal{P}(h)~(h = K, \pi)$ are the probabilities for a track to be assigned to a certain hadron type, based on the $dE/dx$ information measured by the MDC and the time measurement in the TOF.

Photon candidates are reconstructed from isolated showers in the EMC. Each photon candidate is required to have a minimum energy of 25 MeV in the EMC barrel region ($\vert\!\cos\theta\vert<0.8$) or 50 MeV in the end cap region ($0.86<\vert\!\cos\theta\vert<0.92$). To suppress electronic noise and showers unrelated to the event, the difference between the EMC time and the event start time is required to be within~(0,~700) ns. At least four good photon candidates are required for each event.
The $\pi^{0}$ candidates are reconstructed from pairs of photons with invariant masses such that $[M_{\gamma\gamma}-M_{\pi^{0}}]\in[-60,\ 40]$ MeV/$c^2$, where $M_{\pi^{0}}$ is the known $\pi^{0}$ mass taken from the Particle Data Group (PDG)~\cite{ParticleDataGroup:2022pth}. An asymmetrical $\pi^0$ mass window is used because the photon energy deposited in the EMC has a long tail on the low energy side. A one-constraint (1C) kinematic fit is performed on the photon pairs, constraining their invariant masses to the nominal $\pi^{0}$ mass.
The $\chi^{2}_{1\rm C}$ of this kinematic fit is required to be less than 25 to remove fake candidates. At least two reconstructed $\pi^{0}$ candidates per event are further required, where two $\pi^0$ candidates in an event don't share the same photons.

The $\Sc$ and $\ASc$ candidates are built from the proton, anti-proton, and neutral pion candidates.
From all possible combinations, the neutral pion candidates yielding the smallest value of $\sigma_{m}=\sqrt{(M_{p\pi^{0}}-M_{\Sigma})^{2}+(M_{\bar{p}\pi^{0}}-M_{\Sigma})^{2}}$ are assigned to the baryon decay.
Here, $M_{\Sigma}$ is the nominal mass of the $\Sigma$ hyperon~\cite{ParticleDataGroup:2022pth}.
According to the fit to the $M_{p\pi^{0}(\bar{p}\pi^{0})}$ spectrum with a double Gaussian function, the signal events are expected to be within a $M_{p\pi^0(\bar{p}\pi^0)}$ mass range of [1.16, 1.21] GeV$/c^2$.
\vspace{-1.3em}
\begin{figure}[h]
  \centering
  \hspace{-7.2em}
  \includegraphics[width= 9.8cm,height=9.3cm]{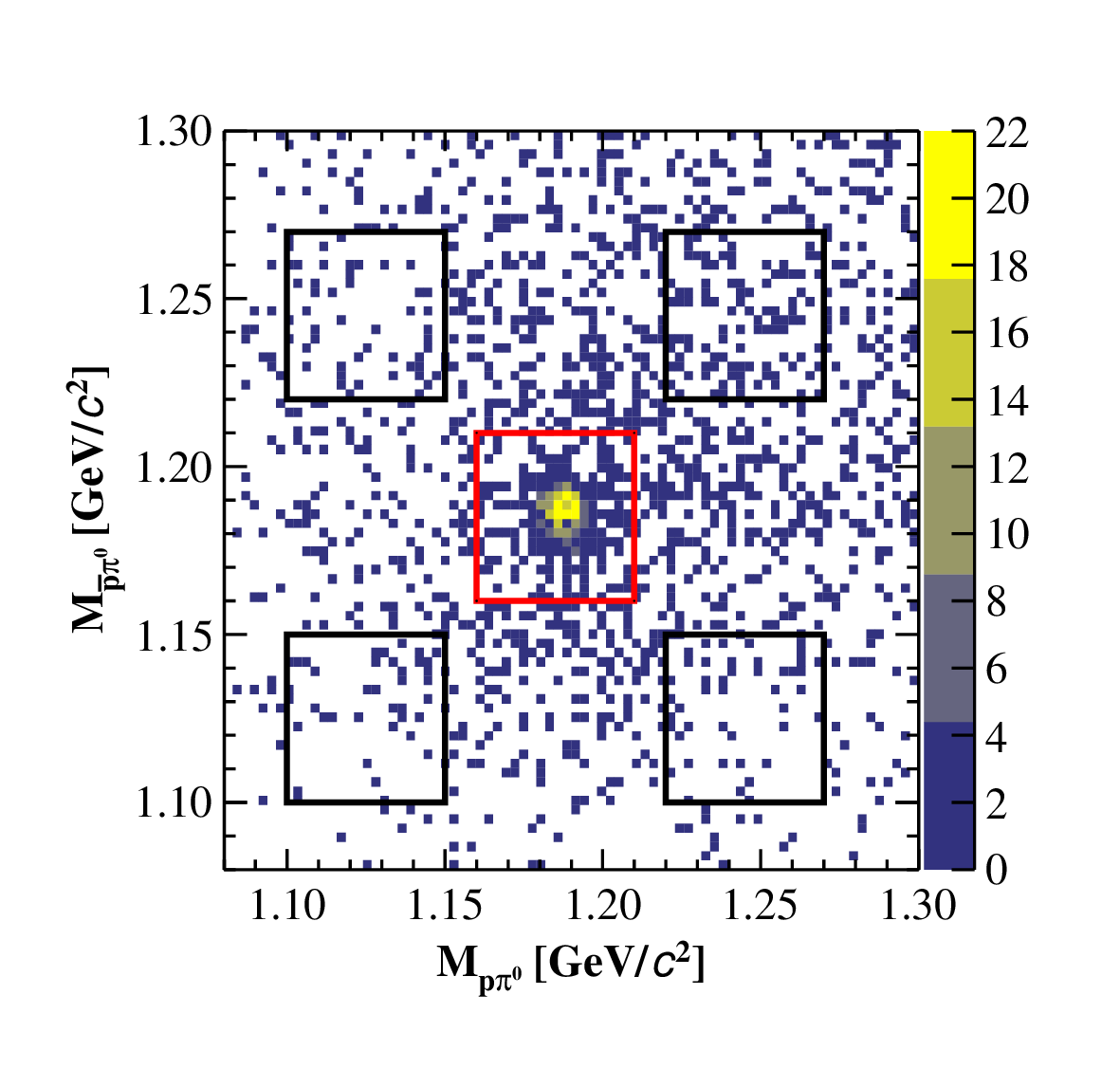}
  \hspace{-8.0em}
  \setlength{\abovecaptionskip}{-1.3em}
  \caption{Distribution of $M_{\bar{p}\pi^{0}}$ versus $M_{p\pi^{0}}$ for the events satisfying the $\eegscsc$ selection criteria from all data sets.
  The red box denotes the signal region and the black ones indicate the sideband regions. }
  \label{fig::2D}
\end{figure}

To further suppress potential background events, the requirements of the ISR photon polar angle $\theta_{\rm miss}<0.25$~radians or $\theta_{\rm miss}>2.90$~radians, and $U_{\rm miss}\in[-0.14, 0.06]$~GeV$/c^2$ are imposed.
Here, $\theta_{\rm miss}$ in the c.m. frame is the opening angle between the momentum of the recoil against the $\Sc\ASc$~system ($P_{\Sc\ASc}^{\rm rec}$) and the beam direction. $U_{\rm miss}=E_{\Sc\ASc}^{\rm rec}-\vert P_{\Sc\ASc}^{\rm rec}\vert$, where $E_{\Sc\ASc}^{\rm rec}$ is the energy of the recoil against the $\Sc\ASc$ system.

\begin{figure}[h]
  \centering
  \hspace{-6.9em}
  \includegraphics[width= 10.4cm,height=7cm]{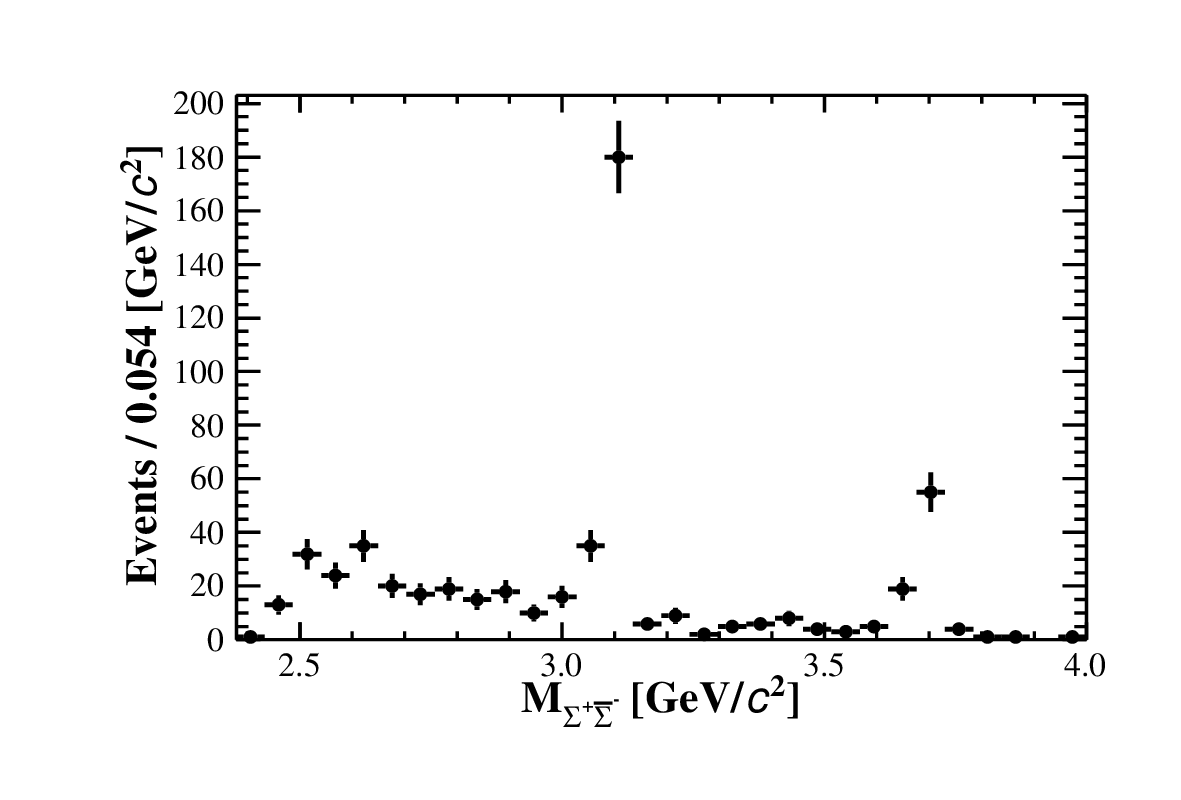}
  \hspace{-6.6em}
  \setlength{\abovecaptionskip}{-1.3em}
  \caption{The $M_{\Sc\ASc}$ distribution for the events satisfying the $\eegscsc$ selection criteria. The black dots with error bars are data combined from all data sets. }
  \label{fig::SCSC}
\end{figure}

Figure~\ref{fig::2D} shows the distribution of $M_{p\pi^{0}}$ versus $M_{\bar{p}\pi^{0}}$ for all data sets combined.
The red box denotes the signal region, which is defined by the mass window discussed above. The black boxes denote the sideband regions used to study and subtract the non-resonant background. For events in the signal region, $M_{\Sc\ASc}$ is plotted in Fig.~\ref{fig::SCSC}, after improving the mass resolution by applying the correction $M_{\Sc\ASc}^{\rm corr}=M_{\Sc\ASc}^{\rm meas}-M_{p\pi^0}-M_{\bar{p}\pi^0}+2M_{\Sigma}$, where $M_{\Sc\ASc}^{\rm meas}$, $M_{p\pi^0}$, and $M_{\bar{p}\pi^0}$ are the measured invariant masses of the four hadrons and the (anti-)proton pion pairs, respectively. Throughout this paper, $M_{\Sc\ASc}$ refers to $M_{\Sc\ASc}^{\rm corr}$.

Potential background sources are investigated by analyzing the inclusive MC samples at $\sqrt{s} =$ 3.773 and 4.178~GeV with the generic event type analysis tool TopoAna~\cite{Zhou:2020ksj}.
After applying the above selection criteria, two background contributions are found to
be insufficiently suppressed: the non-$\Sc\ASc$ channels and the process $\eepiscsc$.
The non-$\Sc\ASc$ background contributions are estimated with the sideband method from the $M_{p\pi^0(\bar{p}\pi^0)}$ distributions, since the corresponding distributions based on the inclusive MC samples after excluding the contributions of channels containing $\Sc\ASc$ pairs do not exhibit significant structures.
The two-dimensional (2D) sideband regions, shown in Fig.~\ref{fig::2D}, are provided in Table~\ref{tab::sidebands}.
The number of the non-\Sc\ASc~background events is obtained by $N_{\rm non-\Sc\ASc}^{\rm bkg}=\frac{1}{4}\Sigma^{4}_{i}N_{\textrm{BG}i}^{\rm data}$, where $i$ runs over the four black boxes shown in Fig.~\ref{fig::2D}.
\begin{table}[htb]
\centering
\caption{Two-dimensional sideband regions on $M_{p\pi^{0}}$ and $M_{\bar{p}\pi^{0}}$ as shown in Fig.~\ref{fig::2D}.}
\label{tab::sidebands}
\begin{tabular}{ccc}
  \hline \hline
  BGi &$M_{p\pi^{0}}$[GeV$/c^2$]&$M_{\bar{p}\pi^{0}}$[GeV$/c^2$] \\ \hline
   BG1 &[1.10, 1.15] &[1.22, 1.27] \\ 
   BG2 &[1.22, 1.27] &[1.22, 1.27] \\
   BG3 &[1.10, 1.15] &[1.10, 1.15] \\
   BG4 &[1.22, 1.27] &[1.10, 1.15] \\
  \hline\hline
\end{tabular}
\end{table}

\vspace{-1.5em}

Events of $\ee\to\pi^{0}\Sc\ASc$ are easily mistaken for signal events if one of the photons of the $\pi^0$s is undetected.
The background contribution to the selected events is estimated by a data-driven approach using the sideband method.
To estimate this contribution, a sample of the $\eepiscsc$ events is selected from data using a similar procedure as described for the ISR production of $\Sc\ASc$, but reconstructing an additional $\pi^0$ candidate instead of a missing photon.
The signal and sideband regions are chosen in the same way as described for the signal process (shown in Fig.~\ref{fig::2D}). The number of events of this sample is calculated by $N_{\pi^0\Sc\ASc}^{\rm data}=N_{\pi^0\Sc\ASc}^{\rm SigReg}-\frac{1}{4}N_{\pi^0\Sc\ASc}^{\rm Side}$, where $N_{\pi^0\Sc\ASc}^{\rm SigReg}$ and $N_{\pi^0\Sc\ASc}^{\rm Side}$ are the numbers of events from the signal and the sideband regions of the $\eepiscsc$ sample, respectively.
The contribution from remaining $\eepiscsc$ background ($N_{\pi^0\Sc\ASc}^{\rm bkg}$) among the signal candidates is determined by:
\begin{equation}
\small{N_{\pi^0\Sc\ASc}^{\rm bkg}=N_{\pi^0\Sc\ASc}^{\rm data}\times\frac{\varepsilon_{\rm bkg}^{\rm MC}}{\varepsilon_{\pi^0\Sc\ASc}^{\rm MC}},}
\end{equation}
where $\varepsilon_{\rm bkg}^{\rm MC}$ and $\varepsilon_{\pi^0\Sc\ASc}^{\rm MC}$ are the detection efficiencies of selecting the $\Sc\ASc$ and $\pi^{0}\Sc\ASc$ candidates, respectively, from the $\eepiscsc$ MC samples.

Figure~\ref{fig::SCSC1} shows the $M_{\Sc\ASc}$ distribution from threshold up to 3.04 GeV$/c^2$ for the events selected from all data sets in Table~\ref{datasamples}. The red and blue histograms indicate the background contributions due to non-$\Sc\ASc$ channels and $\eepiscsc$, respectively.

\section{Cross section of $\eescsc$~and effective FFs}
\label{sec::crosssection}
\begin{figure}[h]
  \centering
  \hspace{-6.9em}
  \includegraphics[width= 10.4cm,height=7cm]{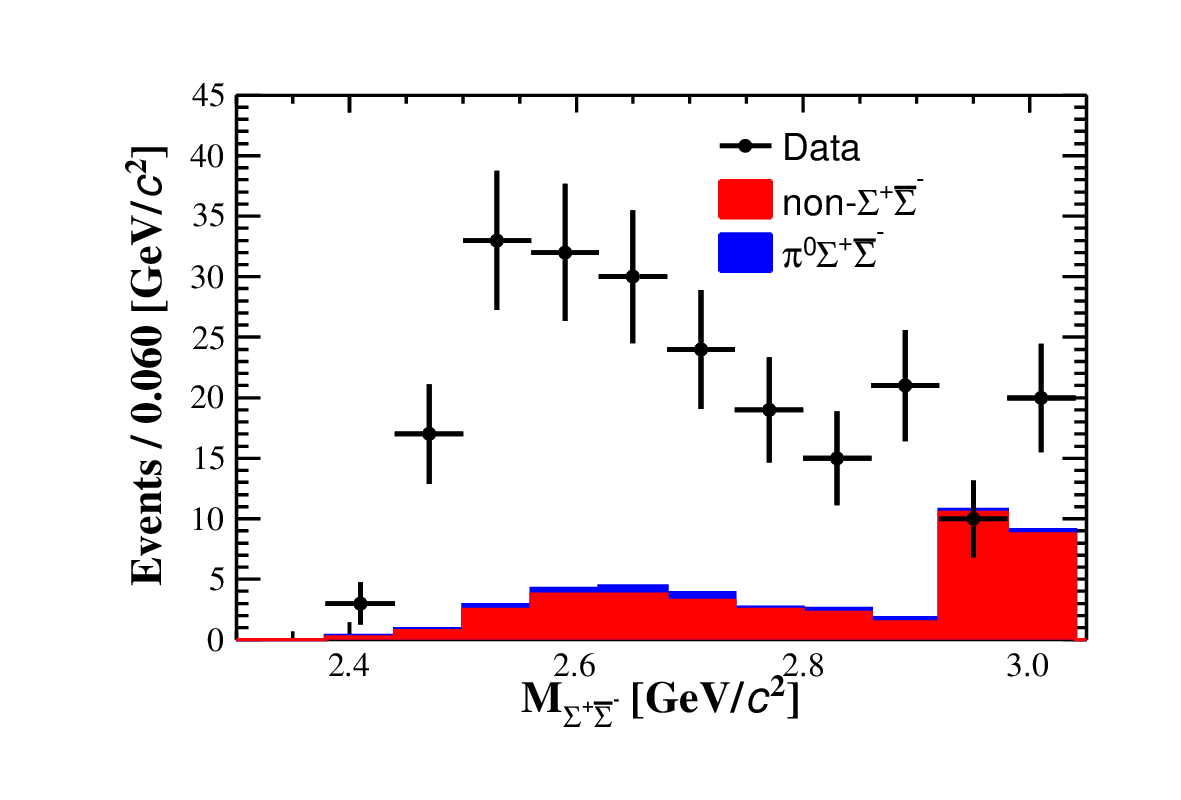}
  \hspace{-6.6em}
  \setlength{\abovecaptionskip}{-1.3em}
  \caption{The $M_{\Sc\ASc}$ distribution for the selected candidates. The black dots with error bars are combined data from all energy points. The red and blue histograms represent the non-$\Sc\ASc$ and $\eepiscsc$ background events, respectively. }
  \label{fig::SCSC1}
\end{figure}

The cross section for the process \eescsc~is calculated from the $M_{\Sc\ASc}$ distribution for each data set by
\begin{equation} \label{eq:XS}
\small{\sigma_{\Sc\ASc}(M_{\Sc\ASc})=
\frac{(dN^{\rm sig}/dM_{\Sc\ASc})}{\varepsilon(\mathcal{B}(\Sigma))^{2}(\mathcal{B}(\pi^{0}))^{2}(d\mathcal{L}_{\rm int}/dM_{\Sc\ASc})},}
\end{equation}
where $\mathcal{B}(\Sigma)$ = ($51.57\pm 0.30$)\% and $\mathcal{B}(\pi^{0})$ = ($99.823\pm0.034$)\% are the branching fractions of $\Sc/\ASc\to p\pi^{0}/\bar{p}\pi^{0}$ and $\pi^{0}\to\gamma\gamma$~\cite{ParticleDataGroup:2022pth}, respectively. $\mathcal{L}_{\rm int}$ is the integrated luminosity, as listed in Table~\ref{datasamples}.
The effective ISR luminosity $(d\mathcal{L}_{\rm int}/dM_{\Sc\ASc})$ $= W(s,x)\mathcal{L}_{\rm int}$ is  calculated by Eq.~\eqref{corr_ISRfact}.
$\varepsilon$ is the detection efficiency determined using the signal MC samples as a function of $M_{\Sc\ASc}$, and combined as the average value weighted by the corresponding effective ISR luminosity. The $(dN^{\rm sig}/dM_{\Sc\ASc})$ is obtained from the $M_{\Sc\ASc}$ distribution of data after subtracting background. The signal yields are extracted by counting the number of observed events in the signal region as shown in Fig.~\ref{fig::SCSC1}.
\vspace{0.1em}

The spectrum of $M_{\Sc\ASc}$ is divided into eleven mass intervals between the $\Sc\ASc$ production threshold and 3.04 GeV$/c^2$, taking into account the mass resolution, which is smaller than fifth of the bin size.
Thus, the spectrum is not unfolded for detector resolution effects.
The measured cross sections and the $\Sigma$ effective FFs calculated by Eqs.~\eqref{eq:a00} and~\eqref{eq:a11} are listed in Table~\ref{tab::crosssection_comb}. The $\overline{\varepsilon}$ and $\mathcal{L}_{\rm eff}$ are the average detection efficiency of all data sets weighted by the effective ISR luminosity and the total effective ISR luminosity, respectively. For the $M_{\Sc\ASc}$ intervals of [2.379, 2.44] and [2.92, 2.98] GeV$/c^2$ where fewer than ten signals are found, the upper limits of the signal yield at 90\% confidence level are calculated using the profile likelihood method~\cite{Lundberg:2009iu}.
\begin{table*}[htb]
\centering
\caption{ The cross sections ($\sigma_{\Sc\ASc}$) and the effective FFs ($|G_{\textrm{eff}}|$) for the process \eescsc~at all data sets. $N^{\rm sig}$ is the total number of signal events. $\overline{\varepsilon}$ is the average detection efficiency of all data sets weighted by the effective ISR luminosity. $\mathcal{L}_{\rm eff}$ is the total effective ISR luminosity.
For $\sigma_{\Sc\ASc}$ and $|G_{\textrm{eff}}|$,  the first and second uncertainties are statistical and systematic, respectively; for $N^{\rm sig}$, the uncertainties are statistical only. The values in brackets for $N^{\rm sig}$, $\sigma_{\Sc\ASc}$ and $|G_{\textrm{eff}}|$ correspond to the upper limits at 90\% confidence level.   }
\label{tab::crosssection_comb}
\begin{tabular}{cccccc}
  \hline \hline
$M_{\Sc\ASc}$[GeV$/c^2$]&$N^{\rm sig}$&$\overline{\varepsilon}[\%]$&$\mathcal{L}_{\rm eff}$[pb$^{-1}$]
  &$\sigma_{\Sc\ASc}$[pb]&$|G_{\textrm{eff}}|(\times 10^{-2})$ \\ \hline
2.379-2.44	  & 2.7$^{+1.8}_{-1.9}$($<$6.8)	 & 0.91	  & 15.13	  &  74$^{+50}_{-52}$$\pm$5($<$190)	  & 14.1$^{+4.8}_{-5.0}$$\pm$0.5($<$22.7)\\ 
2.44-2.50	  & 16$\pm$4	 & 2.07	  & 15.77	  & 190$\pm$50$\pm$20	  & 18.2$\pm$2.4$\pm$1.0\\ 
2.50-2.56	  & 30$\pm$6	 & 3.69	  & 16.82	  & 187$\pm$37$\pm$19	  & 16.6$\pm$1.7$\pm$0.8\\ 
2.56-2.62	  & 28$\pm$6	 & 5.33	  & 17.96	  & 112$\pm$24$\pm$11	  & 12.3$\pm$1.3$\pm$0.6\\ 
2.62-2.68	  & 26$\pm$6	 & 6.49	  & 19.22	  & 79$\pm$18$\pm$8	  & 10.2$\pm$1.2$\pm$0.5\\ 
2.68-2.74	  & 20$\pm$5	 & 7.24	  & 20.62	  & 52$\pm$14$\pm$4	  & 8.1$\pm$1.1$\pm$0.3\\ 
2.74-2.80	  & 16$\pm$5	 & 7.85	  & 22.16	  & 36$\pm$10$\pm$3	  & 6.7$\pm$1.0$\pm$0.3\\ 
2.80-2.86	  & 13$\pm$4	 & 8.19	  & 23.90	  & 26$\pm$8$\pm$2	  & 5.5$\pm$0.9$\pm$0.2\\ 
2.86-2.92	  & 19$\pm$5	 & 8.62	  & 25.83	  & 33$\pm$8$\pm$2	  & 6.5$\pm$0.8$\pm$0.2\\ 
2.92-2.98	  & $-0.7^{+4.5}_{-4.6}$($<$7.7)	 & 8.96	  & 28.01	   & $-1.1^{+6.8}_{-7.1}$$\pm$0.1($<$11.7)	  & $-1.2^{+3.6}_{-3.8}$$\pm$0.1($<$3.9)\\ 
2.98-3.04	  & 11$\pm$6	 & 9.23	  & 30.50	  & 15$\pm$8$\pm$1	  & 4.4$\pm$1.1$\pm$0.1\\ \hline
\hline
\end{tabular}
\end{table*}

\section{systematic uncertainties}
Several sources of systematic uncertainty are considered in the cross section measurement, including the tracking and PID efficiencies of charged tracks, the $\pi^0$ efficiency correction, the $\Sigma$ mass window, the $U_{\rm miss}$ and $\theta_{\rm miss}$ requirements, the background estimation, the angular distribution, the luminosity, and the branching fractions of intermediate states. The individual contributions are discussed below.
\begin{enumerate}
\item \textbf{Tracking and PID efficiencies:}
The tracking and PID efficiency differences between data and MC simulation for the proton and anti-proton have been studied in different bins of transverse momentum and polar angle from the control samples of $J/\psi\to p\bar{p}\pi^{+}\pi^{-}$ and $\psi(3686)\to p\bar{p}\pi^{+}\pi^{-}$~\cite{BESIII:2021wkr}. The differences averaged over the transverse momentum and polar angle of $p$ or $\bar{p}$ of the signal MC samples are taken as the correction factors to calculate the nominal efficiencies. The systematic uncertainty is obtained by summing their relative uncertainties in different bins quadratically and is assigned to be 1.6\% for each $M_{\Sc\ASc}$ interval.

\item \textbf{\bm{$\pi^{0}$} reconstruction:}
Based on the control samples of $\psi(3686)\to\pi^{0}\pi^{0}J/\psi, J/\psi\to l^+l^-$ and $\ee\to\omega\pi^{0}$, the efficiency differences between data and MC simulation are determined as a function of the momentum, $\Delta\varepsilon_{\pi^0}(p)= (0.06-2.41p-\sqrt{0.76p^{2}+1.15+0.39p})\%$~\cite{BESIII:2021wkr}. The systematic uncertainty is obtained by weighting the relative uncertainties according to the momentum distribution in each mass bin,
\begin{equation} \label{eq:MC}
~~~~~~\Delta\varepsilon_{\pi^{0}}^{\rm rec }(p)=\frac{n_1}{N}\Delta\varepsilon_{\pi^{0}}(p_1)+\frac{n_2}{N}\Delta\varepsilon_{\pi^{0}}(p_2)+...,~~~
\end{equation}
where $n_i$ is the number of $\pi^{0}$ candidates in the $i$-th bin and $N$ is the total number of $\pi^{0}$ candidates, both in the signal MC sample. The systematic uncertainties of the $\pi^0$ reconstruction $\Sc\to p\pi^{0}$ and $\ASc\to\bar{p}\pi^{0}$ are both 1.65\%. Therefore, the total systematic uncertainty due to the $\pi^{0}$ reconstruction for \eescsc~ is 3.3\%.

\item \textbf{\bm{$\Sigma$}~mass window and \bm{$U_{\rm miss}$} requirement:}
The uncertainties due to the $\Sigma$~mass window and the $U_{\rm miss}$ requirement are estimated by studying the $\psi(3686)\to\gamma\chi_{c0}, \chi_{c0}\to\Sc\ASc$ decay. The difference in the $\Sigma$~mass window (the $U_{\rm miss}$ requirement) between data and MC simulation which is 1.0(1.4)\%, is taken as the systematic uncertainty.

\item \textbf{\bm{$\theta_{\rm miss}$} requirement:}
The uncertainty due to the $\theta_{\rm miss}$ requirement is estimated by studying $\ee\to\gamma^{\rm ISR}J/\psi, J/\psi\to\Sc\ASc$ decay. The difference in the $\theta_{\rm miss}$ requirement between data and MC simulation which is 2.8\%, is taken as the systematic uncertainty.

\item \textbf{Background estimation:}
To estimate the uncertainties on the number of the non-\Sc\ASc~background events, the 2D sideband regions are changed from [1.10,1.15] and [1.22,1.27] GeV$/c^2$ to [1.095,1.145] and [1.225,1.275] GeV$/c^2$.
The resulting differences to the nominal result are taken as the systematic uncertainties.
The uncertainties of the $\eepiscsc$ background events are estimated by also changing the 2D sideband regions in the event selection, and are included in each $M_{\Sc\ASc}$ interval. The total uncertainty of the background estimation is determined by the average of the uncertainty of all $M_{\Sc\ASc}$ intervals. Thus, the total systematic uncertainty on the background estimation for each \Sc\ASc~mass interval is assigned as 2.5\% at $\sqrt{s}=$ 3.773 GeV and 1.1\% at the other energy points.

\item \textbf{Angular distribution:}
In this analysis, the signal MC samples are generated according to an homogeneous and isotropic phase space population, and the angular distribution of the \Sc\ASc~pair, the spin correlation between \Sc~and \ASc, and the polarization of the \Sc(\ASc) decay are not taken into account. To estimate the uncertainty due to these factors, the signal MC samples with an angular amplitude including these effects are generated. The parametrization of the angular amplitude is the same as that in Ref.~\cite{BESIII:2020uqk}, and the corresponding parameters are set to be 0.56 and 0.25 for the $M_{\Sc\ASc}$ internals from the threshold to 2.68~GeV$/c^2$ and from 2.68 to 3.04 GeV$/c^2$, respectively. The relative difference of the detection efficiency to that based on the phase space distribution is assigned as the uncertainty.

\item \textbf{Luminosity:}
The integrated luminosity is measured by using large-angle Bhabha events with an uncertainty of 0.5\% at $\sqrt{s}=$ 3.773 GeV~\cite{BESIII:2015equ} and 1.0\% at $\sqrt{s}=$ 4.128-4.258 GeV~\cite{BESIII:2023ioy,BESIII:2020eyu,BESIII:2015qfd}. Besides, an additional uncertainty of 0.5\% is taken from the radiator function Eq.~\eqref{corr_ISRfact} of Ref.~\cite{Druzhinin:2011qd}. Therefore, the total systematic uncertainty associated with the luminosity for each \Sc\ASc~mass interval is 0.8\% at $\sqrt{s}=$ 3.773 GeV and 1.2\% at the other energy points.

\item \textbf{Quoted branching fractions:}
The branching fractions of $\Sc\to p\pi^{0}$, $\ASc\to \bar{p}\pi^{0}$ and $\pi^{0}\to\gamma\gamma$ are quoted from the PDG~\cite{ParticleDataGroup:2022pth}. The total uncertainty associated with the quoted branching fractions is 1.2\%.

\end{enumerate}

In this analysis, the data sets taken at twelve c.m. energy points are used and they are separated into two groups.
The first group only includes the one at $\sqrt{s}=$ 3.773~GeV and the second group includes the data sets taken at $\sqrt{s}=$ 4.128-4.258 GeV.
The uncertainties of the second group are studied together or obtained from the result at $\sqrt{s}=4.178$~GeV.
The systematic uncertainties of the combined groups are listed in Table~\ref{tab::XSsys} using the two mass intervals with largest statistics as examples.
Uncertainties of the two groups are combined as the average value weighted by the individual detection efficiencies
and effective ISR luminosities.
The weighted average formula is
\begin{eqnarray}\label{eq:aW}
\sigma_{\rm tot}^{2}&=&\sum^{2}_{i=1}\omega_{i}^{2}\sigma_{i}^{2}+\sum_{i,j=1;i\neq{j}}^{2}\rho_{ij}\omega_{i}\omega_{j}\sigma_{i}\sigma_{j},\nonumber
\end{eqnarray}
with
\begin{eqnarray}
\omega_{i}&=&\frac{\varepsilon_{i}(d\mathcal{L}_{\rm int}/dM_{\Sc\ASc})_{i}}{\sum^{2}_{i=1}\varepsilon_{i}(d\mathcal{L}_{\rm int}/dM_{\Sc\ASc})_{i}},
\end{eqnarray}
where $\omega_i$, $\sigma_i$, and $\varepsilon_{i}$ with $i$ = 1, 2 are the weight factor, systematic uncertainty, and detection efficiency for the $i$-th group, $\rho_{ij}$ is the correlation parameter for the $i$-th and $j$-th group.
For the contributions to the systematic uncertainties due to background no correlation is assumed ($\rho_{ij}$ = 0), while full correlation is assumed ($\rho_{ij}$ = 1) for all other contributions.

\begin{table}[htb]
\centering
\caption{The relative systematic uncertainties (in \%) in the measurements of the cross sections for \eescsc~in two $M_{\Sc\ASc}$ intervals for the full data sample. The total uncertainty is obtained by adding all items in quadrature.}
\label{tab::XSsys}
\begin{tabular}{ccc}
  \hline \hline
  Source&2.50-2.56~[GeV$/c^2$]&2.56-2.62~[GeV$/c^2]$ \\ \hline
   Tracking and PID                           &1.6&1.6 \\ 
  $\pi^{0}$ reconstruction                       &3.3&3.3 \\ 
  $U_{\rm miss}$ requirement          &1.4&1.4 \\ 
  $\theta_{\rm miss}$ requirement     &2.8&2.8 \\ 
  $\Sigma$ mass window           &1.0&1.0 \\ 
  Background estimation   	   	         &1.6&1.6 \\ 
  Angular distribution   &	  	8.5  &	8.1  \\ 
  Luminosity                   &0.9&0.9 \\ 
  $\mathcal{B}(\Sigma,\pi^{0})$              &1.2&1.2 \\ \cline{1-3}
  Total  &	   10.0  &	9.6   \\ 
  \hline\hline
\end{tabular}
\end{table}

\section{Branching fractions of $J/\psi\to\Sc\ASc$ and $\psi(3686)\to\Sc\ASc$}
\begin{figure*}[htp]
  \centering
  \hspace{-1.5em}
  \subfigure{\includegraphics[width=0.265\textwidth,height=0.18\textwidth]{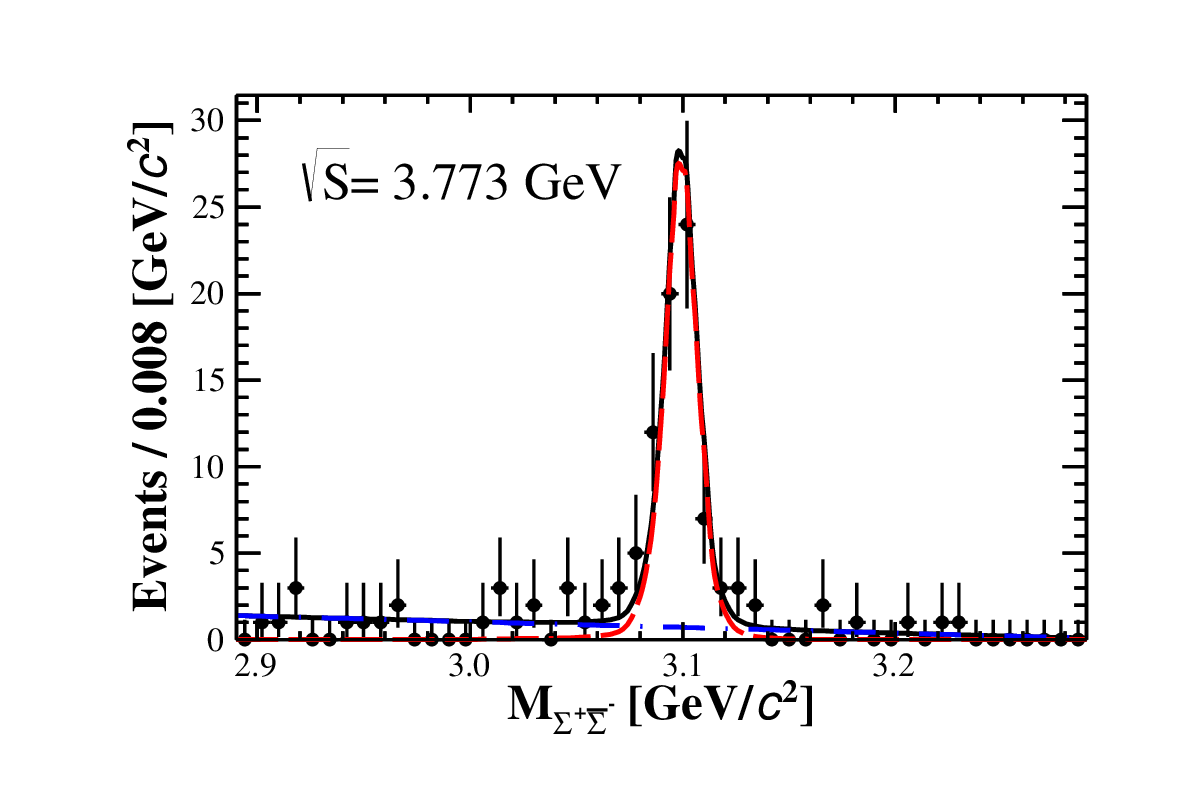}\label{result:jpsi1}}\put(-38,70){ ~(a)}
  \hspace{-1.5em}
  \subfigure{\includegraphics[width=0.265\textwidth,height=0.18\textwidth]{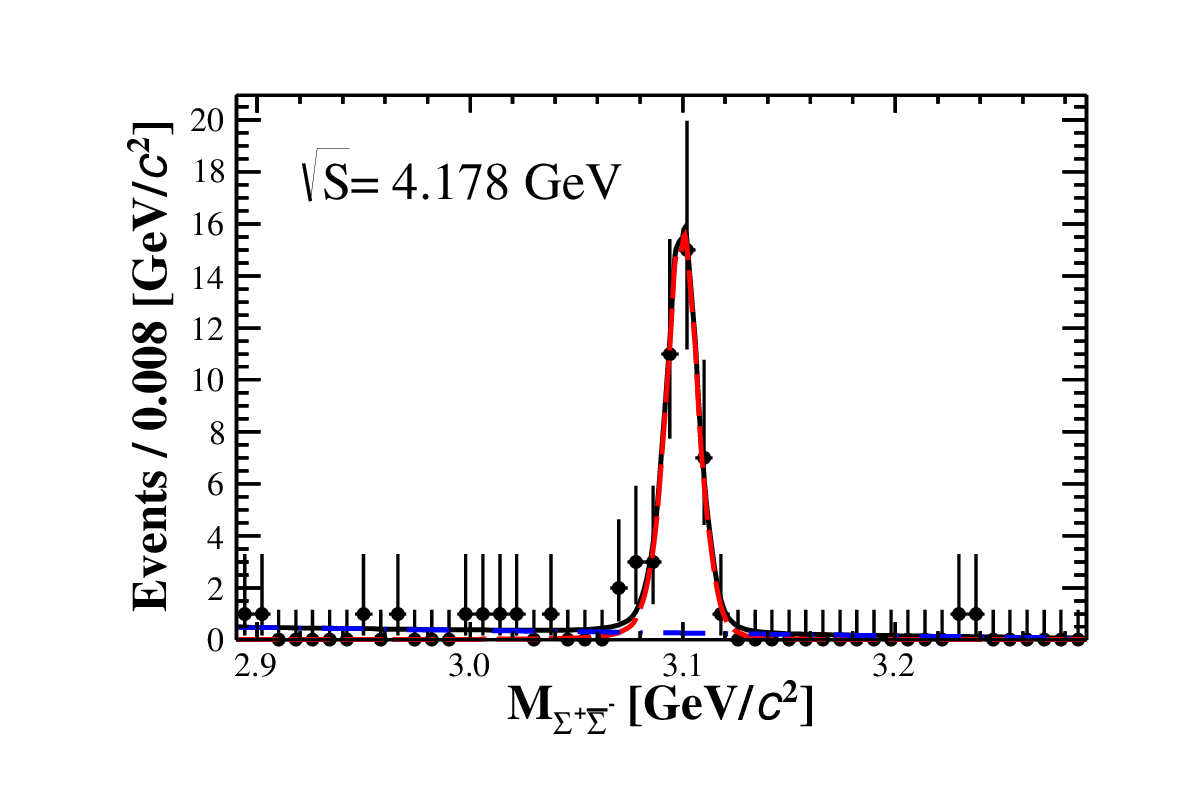}\label{result:jpsi2}}\put(-38,70){ ~(b)}
  \hspace{-1.5em}
  \subfigure{\includegraphics[width=0.265\textwidth,height=0.18\textwidth]{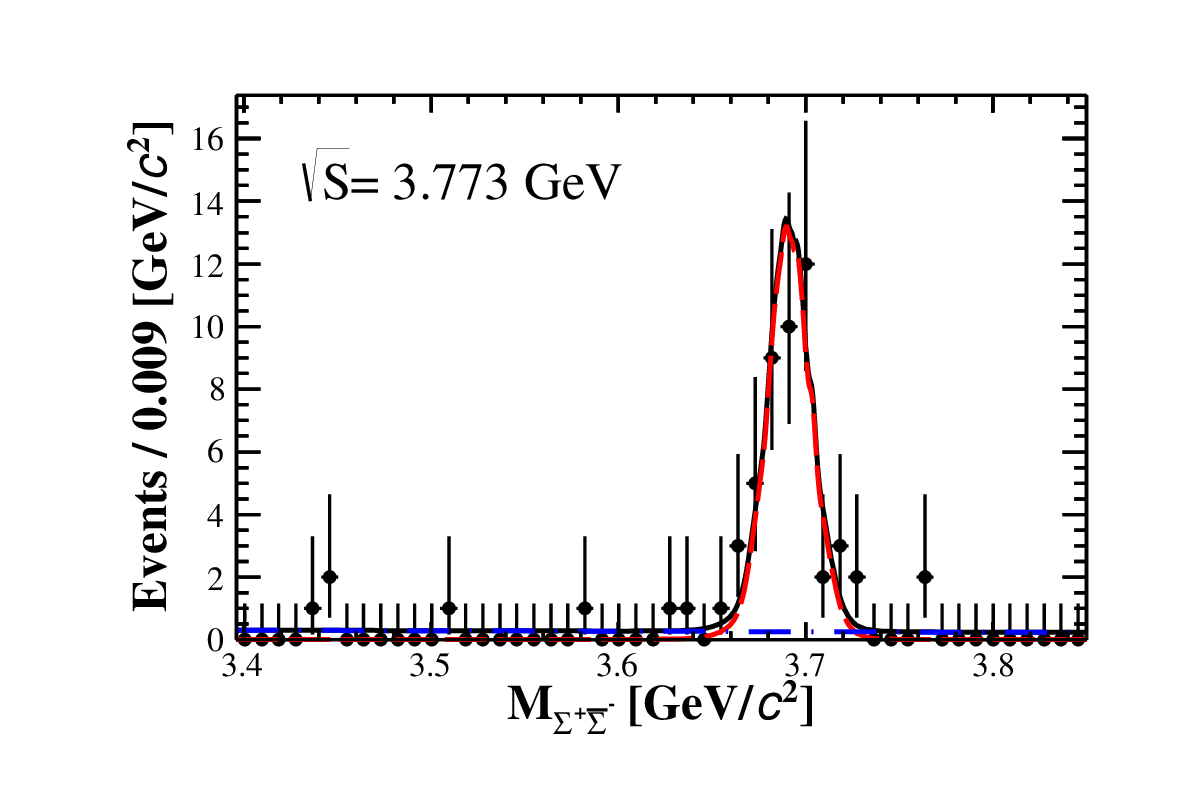}\label{result:psip1}}\put(-38,70){ ~(c)}
  \hspace{-1.5em}
  \subfigure{\includegraphics[width=0.265\textwidth,height=0.18\textwidth]{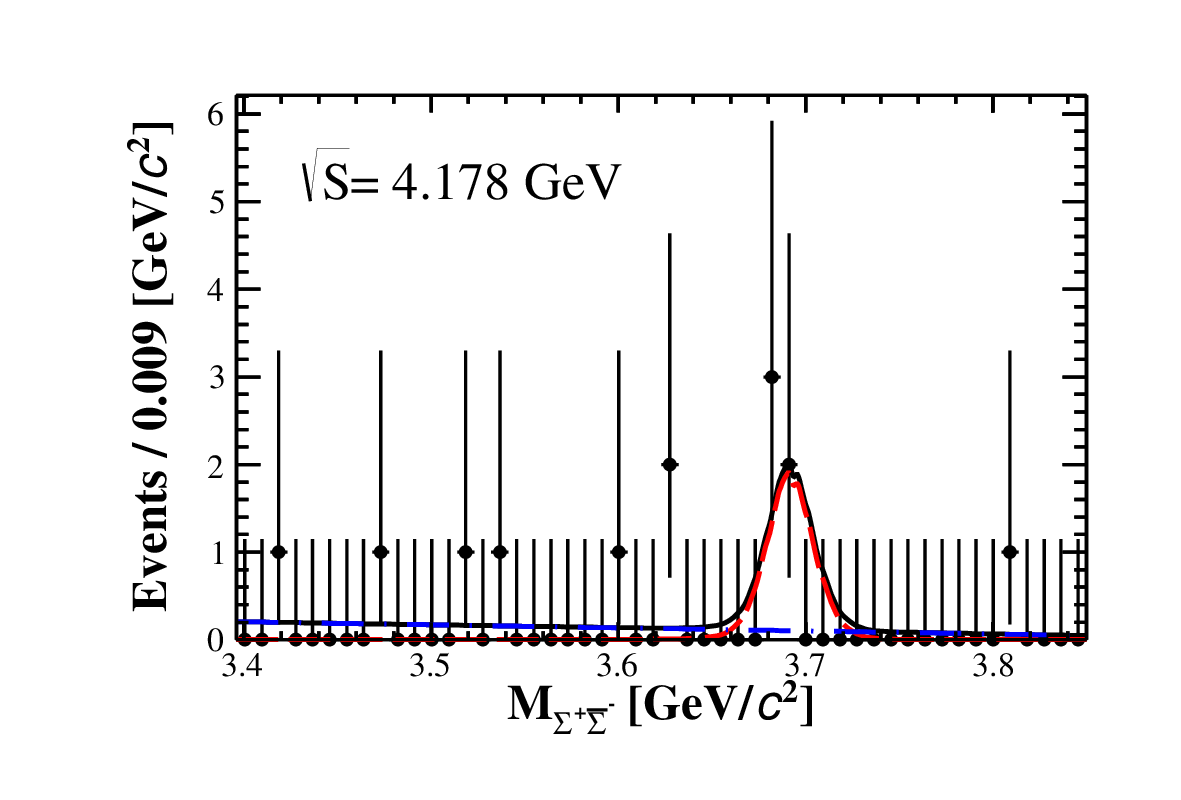}\label{result:psip2}}\put(-38,70){ ~(d)}
  \setlength{\abovecaptionskip}{-0.1em}
  \caption{Simultaneous fit (black curve) with the MC simulated shape (red dashed curve) for the resonance of (a) and (b) $J/\psi$ or (c) and (d) $\psi(3686)$ and a linear function (blue dashed curve) for the background of the $M_{\Sc\ASc}$ spectra at $\sqrt{s}=3.773$ and $4.178$~GeV. The black dots with error bars are data. \label{fig::states}}
\end{figure*}

As shown in Fig.~\ref{fig::SCSC}, masses from the $\Sigma$ pair threshold up to 4.0 GeV$/c^2$ can be studied, including the narrow charmonium vector resonances $J/\psi$ and $\psi(3686)$.
Thus, it is possible to study the decays of these resonances into pairs of hyperons and to determine the branching fractions $\mathcal{B}(J/\psi, \psi(3686)\to\Sc\ASc)$ by using the data sets taken at $\sqrt s=3.773$ and 4.178 GeV with the ISR method.
After integrating over the ISR photon polar angle, the cross section for ISR production of a narrow vector meson resonance 
decaying into the final state $\Sc\ASc$ can be calculated by~\cite{BESIII:1999res}:
\begin{equation} \label{XS_resonance}
\small{\sigma(s)=\frac{12\pi^{2}\Gamma(V\to e^{+}e^{-})\mathcal{B}(V\to\Sc\ASc)}{M_{V}s}W(s, x_{0}),}
\end{equation}
where $M_{V}$ and $\Gamma(V\to\EE)$ are the mass and electronic width of the vector meson $V$. Here, $V$ are the $J/\psi$ and $\psi(3686)$ with $\Gamma(J/\psi\to\ee)=\Gamma(J/\psi)\cdot\mathcal{B}(J/\psi\to\ee)= (5.529\pm0.106)$~keV and $\Gamma(\psi(3686)\to\ee)=\Gamma(\psi(3686))\cdot\mathcal{B}(\psi(3686)\to\ee)= (2.331\pm0.063)$~keV~\cite{ParticleDataGroup:2022pth}, respectively. $\mathcal{B}(V\to \Sc\ASc)$ is the branching fraction of $V\to\Sc\ASc$.
$W(s, x_{0})$ is calculated using Eq.~\eqref{corr_ISRfact} with $x_{0}=1-M_{V}^{2}/s$. If the cross section is measured, the branching fraction can be calculated by Eq.~\eqref{XS_resonance}. The cross section can also be written as:
\begin{equation}\label{XS}
\small{\sigma(s)=\frac{N^{\rm sig}_{V}}{\mathcal{L}_{\rm int}\varepsilon_{V}(\mathcal{B}(\Sigma))^{2}(\mathcal{B}(\pi^{0}))^{2}},}
\end{equation}
where $N_{V}^{\rm sig}$ is the number of $V$ events. The $\varepsilon_{V}$ is the detection efficiency of $\ee\to\gamma^{\rm ISR}V\to\gamma^{\rm ISR}\Sc\ASc$ determined from the signal MC samples. The MC samples are generated with $\Sigma$ angular distributions described as $1+\eta\cos^{2}\theta_{\Sigma}$ with $\eta=-0.508$ for $J/\psi$ and $\eta=0.682$ for $\psi(3686)$~\cite{BESIII:2020fqg}.
The remaining parameters in Eq.~\eqref{XS} are consistent with those defined above.
To extract $N_{V}^{\rm sig}$, using $\mathcal{B}(V\to\Sc\ASc)$ as a shared parameter, a simultaneous fit to the $M_{\Sc\ASc}$ distributions at $\sqrt s=3.773$ and 4.178 GeV is performed. The MC simulated shape is used to describe the resonance and a linear function for the background and the continuum contribution.
The fit results are shown in Figs.~\ref{fig::states}(a) and~\ref{fig::states}(b) for $J/\psi$ and Figs.~\ref{fig::states}(c) and~\ref{fig::states}(d) for $\psi(3686)$.

The systematic uncertainties in the measurements of $\mathcal{B}(J/\psi, \psi(3686)\to\Sc\ASc)$ include tracking and PID efficiencies of charged tracks, $\pi^{0}$ reconstruction, the $U_{\rm miss}$ requirement, the $\theta_{\rm miss}$ requirement, the $\Sigma$ mass window, the luminosity, and the branching fractions of intermediate states. They are assigned in the same way as for the cross section measurement. In addition, the uncertainty due to the MC model is considered by changing the generator model for the decays of $J/\psi$ or $\psi(3686)$ from \textsc{HELAMP} to \textsc{AngSam}~\cite{Ping:2008zz}.
The uncertainty of the fit region is determined by changing the fit region from [2.89, 3.29] GeV/$c^2$ to a wider [2.79, 3.30] GeV/$c^2$ and a narrower interval [2.94, 3.24] GeV/$c^2$.
The uncertainty of the signal model is estimated by additionally convolving the signal shape with a Gaussian distribution to account for the possible differences between data and MC simulation. The uncertainty of the background model in the fit is estimated by changing the model from a linear function to a constant, which is found to be negligible. The systematic uncertainties of the measurements of the branching fractions of $J/\psi\to\Sc\ASc$ and $\psi(3686)\to\Sc\ASc$ are listed in Table~\ref{tab::BFsys}.
\begin{table}[htb]
\centering
\caption{The relative systematic uncertainties (in \%) in the measurements of the branching fractions of $J/\psi\to\Sc\ASc$ and $\psi(3686)\to\Sc\ASc$. The total uncertainty is obtained by adding all items in quadrature.}
\label{tab::BFsys}
\begin{tabular}{ccc}
  \hline \hline
Source&$J/\psi$&$\psi(3686)$\\ \hline
  Tracking and PID                                                    &1.6&1.5\\
  $\pi^{0}$ reconstruction                                          &3.3&3.3\\
  $U_{\rm miss}$ requirement                                 &1.4&1.4\\
  $\theta_{\rm miss}$ requirement                            &2.8&2.8\\
  $\Sigma$~mass window                              &1.0&1.0\\
  MC model                                                  &0.5&3.1\\
  Luminosity                                      &0.9&0.9\\
  $\mathcal{B}(\Sigma,\pi^{0})$                                 &1.2&1.2\\
  Fit range                                                    &1.7&0.8\\
  Signal model of the fit                                       &2.7&4.4\\ \hline
  Total                                                         &6.0&7.4\\ \hline
  \hline
\end{tabular}
\end{table}

The final results for $\mathcal{B}(J/\psi, \psi(3686)\to\Sc\ASc)$ with Eqs.~\eqref{XS_resonance} and \eqref{XS} are listed in Table~\ref{tab::BF_states}.
They are consistent with the previous results by BESIII with the data sets taken at the $J/\psi$ or $\psi(3686)$ resonance peak~\cite{BESIII:2021wkr}, 
showing the reliability of the method of determining the $\Sigma$ EMFFs.

\begin{table}[h]
\centering
\caption{ The branching fractions of $J/\psi\to\Sc\ASc$ and $\psi(3686)\to\Sc\ASc$ (in $10^{-4}$), where the first and second uncertainties
are statistical and systematic, respectively. }
\label{tab::BF_states}
\begin{tabular}{ccc}
  \hline \hline
  Decay                & This work                   & Previous results~\cite{BESIII:2021wkr} \\ \hline
  $J/\psi\to\Sc\ASc$     &$8.88\pm0.90\pm0.53$                          &$10.61\pm0.04\pm0.36$\\  
  $\psi(3686)\to\Sc\ASc$ &$2.51\pm0.40\pm0.19$                          &$2.52\pm0.04\pm0.09$\\  \hline
  \hline
\end{tabular}
\end{table}

\section{The line-shape analysis}
\begin{figure}[htp]
\centering
\hspace{-6.8em}
\includegraphics[width= 10.6cm,height=7.0cm]{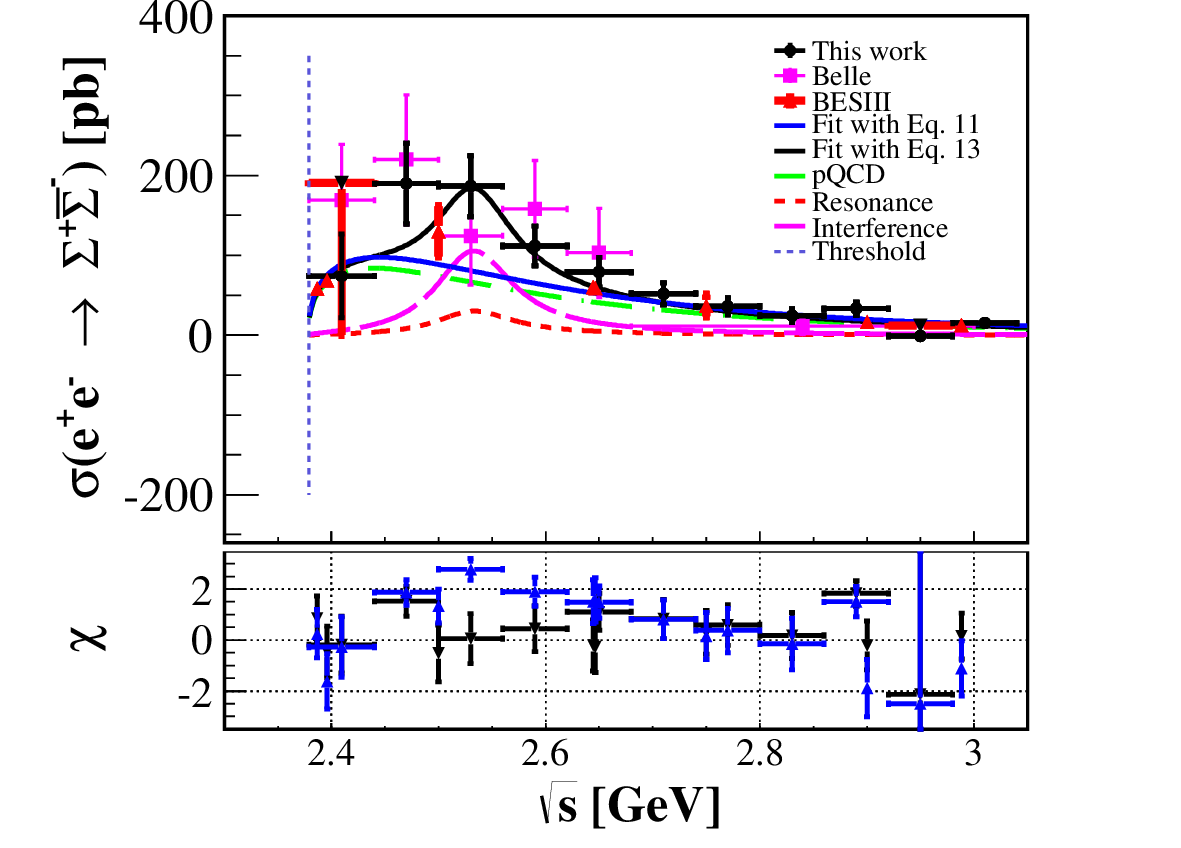}
\hspace{-8.8em}
\setlength{\abovecaptionskip}{0.2em}
\caption{Fit to the cross section of \eescsc. The blue line is the fit result using the perturbative QCD-motivated energy power function, described with Eq.~\eqref{eq:method1}. The black curve represents the total fit result using Eq.~\eqref{eq:method03}. The individual contributions associated with Eq.~\eqref{eq:method03} are indicated by the green curve (pQCD), red curve (Resonance) and purple dashed curve (interference between the pQCD function and the resonance). The vertical dashed line refers to the production threshold for $\eescsc$. The fit quality is presented by $\chi$ distributions and depicted in the bottom panel. The inverted triangle symbol with red line represents its upper limit.%
  \label{fig::Line} }
\end{figure}

The cross sections measured in Section~\ref{sec::crosssection} are consistent with the previous results from BESIII~\cite{BESIII:2020uqk} and Belle~\cite{Belle:2022dvb}, as depicted in Fig.~\ref{fig::Line}.
A search for a threshold effect is made by performing a least chi-square fit to the cross section in this measurement from the production threshold up to 3.04~GeV$/c^2$ and the BESIII scan results~\cite{BESIII:2020uqk} with different functions. The systematic uncertainty is included in the fit with the correlated and uncorrelated parts considered separately.

The perturbative QCD-motivated (pQCD)~\cite{Pacetti:2014jai} energy power function is assumed to model the line-shape of \eescsc~production, and the formula is expressed as:
\begin{equation} \label{eq:method1}
\small{
\sigma_{Y\bar{Y}}(s)=\frac{C\beta} {s}\left(1+\frac{2M_{Y}^{2}}{s}\right) \frac{c_0}{(s-c_1)^4[\pi^2 + {\rm ln}^2(s/\Lambda^2_{\rm QCD})]^2},
}
\end{equation}
where $c_0$ is the normalization parameter, $c_1$ is the contribution of resonant states, $\Lambda_{\rm QCD}$ is the QCD scale fixed to 0.3 GeV and all parameters and variables are consistent with those defined for Eq.~\eqref{eq:a00}. The fit result is shown as the solid blue line in Fig.~\ref{fig::Line} and the parameters are listed in Table~\ref{tab::parameter}, with the fit quality $\chi^2/n\rm dof=31.9/17$, where the number of degrees of freedom ($n\rm dof$) is calculated by subtracting the number of free parameters in the fit from the total number of $M_{\Sc\ASc}$ intervals. The bottom panel of Fig.~\ref{fig::Line} present the $\chi$ distributions defined by $\frac{\sigma_{\Sc\ASc}-\sigma_i}{\Delta\sigma_{\Sc\ASc}}$, where the $\sigma_{\Sc\ASc}$ and $\sigma_i$ are the measured and fitted cross sections at each \Sc\ASc~invariant mass interval, respectively, and $\Delta\sigma_{\Sc\ASc}$ corresponds to the error of the measured value which is counted as the quadratic sum of the statistical and systematic uncertainties.

\vspace{1.5em}

The Fano-type FF includes the interference between several resonances and the continuum contribution~\cite{Fano:1961zz}.
The amplitude of the resonance can be written in terms of Breit-Wigner parametrizations,
\begin{equation} \label{eq:method02}
\small{BW(s)=\frac{\sqrt{12\pi\Gamma^{ee}\mathcal{B}\Gamma}}{s^2-M^{2}+iM\Gamma},}
\end{equation}
where $M$ and $\Gamma$ are the mass and width of the resonance, $\Gamma^{ee}$ and $\mathcal{B}$ are the corresponding electronic partial width and branching fraction, respectively.

Assuming the pQCD-motivated power function to describe the continuum contribution, the line-shape of the cross section for \eescsc~can be modeled by the coherent sum of Eqs.~\eqref{eq:method1} and \eqref{eq:method02} to account for the possible interference of a vector meson and the continuum production:
\begin{equation} \label{eq:method03}
\small{
\begin{aligned}
\sigma_{Y\bar{Y}}(s)=\Bigl|\sqrt{\frac{C\beta}{s}(1+\frac{2M_{Y}^2}{s}) \frac{c_0}{(s-c_1)^4[\pi^2 + {\rm ln}^2(s/\Lambda^2_{\rm QCD})]^2}}
\\+e^{i\phi}BW(s)\sqrt{\frac{P(s)}{P(M)}}\Bigl|^{2}.~~~~~~~~~~~~~~~~~~~~~~~~~~~~~~~~~~
\end{aligned}
}
\end{equation}
Here, $\phi$ is a relative phase between the $BW(s)$ function and the pQCD energy power function, and $P(s)$ is the two-body phase space factor.
The fit result assuming the line-shape as described in Eq.~\eqref{eq:method03} is shown as the solid black line in Fig.~\ref{fig::Line} and the parameters are listed in Table~\ref{tab::parameter}, with the fit quality $\chi^2/n\rm dof$ being 12.6/13.

\begin{table}[h]
\centering
\caption{ The parameters obtained from the fit to the cross section line-shape. The method 1 uses Eq.~\eqref{eq:method1}, while the method 2 uses Eq.~\eqref{eq:method03}. }
\label{tab::parameter}
\begin{tabular}{ccc}
  \hline
  \hline
  Method &Parameter           &\eescsc            \\ \hline
  \multirow{2}{*}{Method1}&$c_0\times10^{4}$[$\rm pb^{-1}.GeV^{10}$]    &3.61$\pm$0.47       \\ 
                          &$c_1$[GeV]                &1.79$\pm$0.03              \\ \hline
  \multirow{6}{*}{Method2}&$c_0\times10^{4}$[$\rm pb^{-1}.GeV^{10}$]    &2.18$\pm$2.16       \\ 
                          &$c_1$[GeV]                &1.85$\pm$0.17              \\ 
                          &$\Gamma^{ee}\mathcal{B}$  &0.45$\pm$0.26\\ 
                          &$M$[GeV$/c^2$]                  &2.53$\pm$0.04\\ 
                          &$\Gamma$[MeV$/c^2$]             &88.67$\pm$42.09\\ 
                          &$\phi$[rad]                    &1.42$\pm$1.02\\ \hline
  \hline
\end{tabular}
\end{table}

According to the difference of $\chi^{2}/n\rm dof$ between the results, the statistical significance for the model indicating the presence of a resonance state near 2.5 GeV is estimated to be 3.4$\sigma$, including both statistical and systematic uncertainties. Thereby, there may be resonant structures at 2.5 GeV.

\begin{figure}[htp]
\centering
\hspace{-6.9em}
\includegraphics[width= 10.4cm,height=7cm]{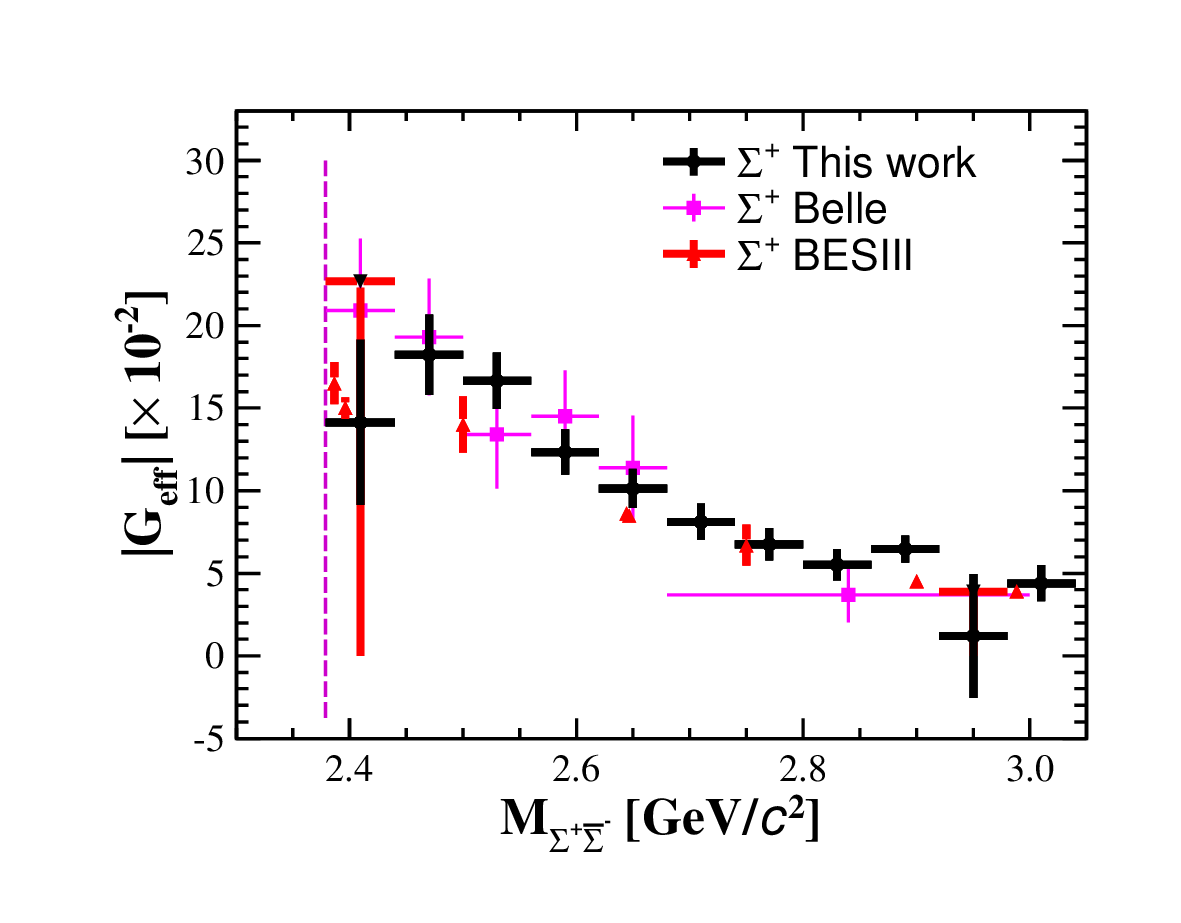}
\hspace{-6.3em}
\setlength{\abovecaptionskip}{-0.5em}
\caption{ Comparison of the $\Sigma$ effective FFs obtained in this work with the previous measurements of BESIII~\cite{BESIII:2020uqk} and Belle~\cite{Belle:2022dvb}. The inverted triangle symbol with red line represents its upper limit. The vertical dashed line refers to the production threshold for $\eescsc$.
  \label{fig::Line1} }
\end{figure}

Fig.~\ref{fig::Line1} shows a comparison of the measured $\Sigma$ effective FFs with previous measurements of $\Sigma^{+}$ at BESIII~\cite{BESIII:2020uqk} and Belle~\cite{Belle:2022dvb}. A prediction for the non-resonant cross section of $\eescsc$ at the $J/\psi$ mass~\cite{BaldiniFerroli:2019abd}, based on an effective Lagrangian density, is consistent with our result when extrapolated to $\sqrt{s} =$ 3.097~GeV using Eq.~\eqref{eq:method1}.

\vspace{-1.0em}
\section{Summary}
Using the untagged ISR technique, where the radiative photon is not detected, the cross section of
\eescsc is measured and the effective FF of $\Sigma$ is determined from threshold to 3.04~GeV$/c^2$.
A data set corresponding to a total integrated luminosity of 12.0 fb$^{-1}$, collected at twelve c.m. energies with the BESIII detector at the BEPCII collider, is analyzed.
The results are consistent with the previous measurements from BESIII~\cite{BESIII:2020uqk} and Belle~\cite{Belle:2022dvb}.
It should be noted that the width of the lowest \Sc\ASc~mass interval in this work is about 30.5~MeV above the threshold, and the value in the lowest \Sc\ASc~invariant mass interval (2.379-2.44 GeV$/c^2$) is definitely closer to the BESIII scan results~\cite{BESIII:2020uqk} than to the Belle ISR measurement~\cite{Belle:2022dvb}. Our results also provide experimental inputs to test various theoretical models, such as $Y\bar{Y}$ potential and diquark correlation models~\cite{Anselmino:1992vg, Jaffe:2003sg, Jaffe:2004ph}.
Furthermore, combining the data sets taken at $\sqrt{s}=$ 3.773 and 4.178 GeV, the branching fractions of $J/\psi\to\Sc\ASc$ and $\psi(3686)\to\Sc\ASc$ are measured. The obtained results are consistent with the previous measurements of BESIII~\cite{BESIII:2021wkr}.
In the near future, BESIII will finish collecting data at $\sqrt{s}=$3.773~GeV with a total integrated luminosity of 20~fb$^{-1}$~\cite{BESIII:2020book}, to allow a more precise ISR measurement.

\vspace{-1.0em}
\section{acknowledgements}
The BESIII Collaboration thanks the staff of BEPCII and the IHEP computing center for their strong support. This work is supported in part by National Key R\&D Program of China under Contracts No. 2020YFA0406300 and No. 2020YFA0406400; National Natural Science Foundation of China (NSFC) under Contracts No. 11635010, No. 11735014, No. 11835012, No. 11935015, No. 11935016, No. 11935018, No. 11961141012, No. 12025502, No. 12035009, No. 12035013, No. 12061131003, No. 12192260, No. 12192261, No. 12192262, No. 12192263, No. 12192264, No. 12192265, No. 12221005, No. 12225509, No. 12235017, No. 12122509 and No. 12005311; the Chinese Academy of Sciences (CAS) Large-Scale Scientific Facility Program; the CAS Center for Excellence in Particle Physics (CCEPP); Joint Large-Scale Scientific Facility Funds of the NSFC and CAS under Contract No. U1832207; CAS Key Research Program of Frontier Sciences under Contracts No. QYZDJ-SSW-SLH003 and No. QYZDJ-SSW-SLH040; 100 Talents Program of CAS; The Institute of Nuclear and Particle Physics (INPAC) and Shanghai Key Laboratory for Particle Physics and Cosmology; European Union's Horizon 2020 research and innovation programme under Marie Sklodowska-Curie grant agreement under Contract No. 894790; German Research Foundation DFG under Contracts No. 455635585, Collaborative Research Center Grant No. CRC 1044, No. FOR5327 and No. GRK 2149; Istituto Nazionale di Fisica Nucleare, Italy; Ministry of Development of Turkey under Contract No. DPT2006K-120470; National Research Foundation of Korea under Contract No. NRF-2022R1A2C1092335; National Science and Technology fund of Mongolia; National Science Research and Innovation Fund (NSRF) via the Program Management Unit for Human Resources \& Institutional Development, Research and Innovation of Thailand under Contract No. B16F640076; Polish National Science Centre under Contract No. 2019/35/O/ST2/02907; The Swedish Research Council; and U. S. Department of Energy under Contract No. DE-FG02-05ER41374.

\end{document}